\documentclass[12pt]{article}

\usepackage[a4paper,left=30mm,right=20mm,top=20mm,bottom=20mm]{geometry}
\usepackage{graphics}
\usepackage{amsmath}
\usepackage{epsfig}
\usepackage{cite}
\newcommand{\be}{\begin{eqnarray}}
\newcommand{\ee}{\end{eqnarray}}
\title{Self-similarity of hadron production
in $pp$ and $AA$ collisions at high energies }

\author{D.A. Artemenkov,G.I. Lykasov,A.I. Malakhov}

\begin{document}
\date{}
\maketitle

\begin{center}

{Joint Institute for Nuclear Research, Dubna 141980, Moscow region, Russia

\vspace{0.5cm}

artemenkov.denis@gmail.com\\
lykasov@jinr.ru\\
malakhov@lhe.jinr.ru
}

\end{center}

\vspace{0.5cm}

\begin{center}

{\bf Abstract }

\end{center} 

We analyze the self-similarity approach applied to study the hadron production in $pp$ and $AA$ 
collisions. 
This approach allows us to describe rather well the
ratio of the proton to anti-proton yields in A-A collisions as a function of the energy at a wide 
range from a few GeV to a few TeV. We suggest a modification of this approach to describe rather 
well the inclusive spectra of hadrons produced in $pp$ collisions at different initial energies    
from the AGS to LHC.

\vspace{1.0cm}

\noindent

\indent

\section{Introduction}
\label{intro}
The description of hadron production using statistical models
has been pioneered several decades ago  by 
E.Fermi \cite{Fermi:1950}, I.Pomeranchuk \cite{Pomeran:1951}, L.D.Landau\cite{Landau:1953}
and  R.Hagedorn \cite{Hagedorn:1965}. 
The transverse momentum  spectrum of particles
produced in hadron-hadron collisions can be presented in a simple form
$\rho_h\sim\exp(-m_{ht}/T)$,
 where $m_{ht}$ is the transverse mass of the hadron $h$ and $T$
is sometimes called the {\it thermal freeze-out} temperature.

As it is well known, the statistical (thermal) models have  been applied successfully to describe hadronic 
yields produced in heavy-ion 
collisions (see, for example, \cite{Wheaton}-\cite{Sinyukov:02}
and references therein). 
The temperature obtained in these analyzes is often referred to as {\it chemical freeze-out} temperature
and is consistently slightly higher than the thermal freeze-out temperature.
At the same time, the source of very fast thermalization is currently unknown and  
alternative or complementary possibilities to explain the thermal spectra are of much interest.

There exists a rich and wide variety of distributions covering a 
large range of applications~\cite{newman,clauset,levy}.
Those having a power law behaviour have attracted considerable attention in physics in 
recent years but there is a a long history 
in other fields such as biology and economics~\cite{mitzenmacher}.

In high energy physics the power law distributions have been applied 
in~\cite{STAR,PHENIX,ALICE,ATLAS,CMS} to the description of transverse momenta  
of secondary particles produced in $pp$ collisions. 
Indeed the available 
range of transverse momenta  has expanded
considerably  with the advent of the Large Hadron Collider (LHC). 
Collider  energies of 7-8 TeV are now available in $pp$ collisions and 
transverse momenta of hundreds of GeV are now  common.
Applications of the Tsallis 
distribution to high energy $e^+ e^-$ annihilation have been considered previously
 in~\cite{bediaga}. A recent review of
power laws in elementary and heavy-ion collisions 
can be found in~\cite{wilkreview}.
The modification of the Tsallis distribution \cite{tsallis1,tsallis2} and its successful application to the analysis of the LHC data 
on the multiple hadron production in $pp$ collisions at the central rapidity region has been done recently in \cite{CLPST:2014}. 

There are other approaches, like the quark gluon string model (QGSM) \cite{ABK:1982,ABK:1999} or the Monte Carlo (MC) versions of string 
model \cite{Lund, Venus}, which are applied to analyze the hadron production mainly at the non central rapidity region. 

The inclusive spectra of hadrons produced in central $pp$ collisions at low and large hadron transverse 
momenta $p_T$  were analyzed within the modified quark gluon string 
model (QGSM) and the perturbative QCD (PQCD) \cite{c12}. There was suggested a contribution of the nonperturbative gluons 
at low transfer momentum squared $Q^2$ \cite{c12,c13}, which results in the satisfactory
description of LHC data on the hadron $p_t$-spectra at the mid-rapidity region and different energies ($\sqrt{s}$).
The description of the energy dependence of these spectra is an advantage in comparison with many theoretical approaches mentioned above.

Almost all theoretical approaches operate the relativistic invariant Mandelstam variables $s,t,u$
to analyze the hadron inclusive spectra in the  mid-rapidity region. As usual, the spectra are presented in the factorized forms of 
two functions dependent of $t$ or $p_t^2$ and $s$.
However, there is another approach to analyze multiple hadron production in $pp$ and $AA$ collisions at high energies,
which operates the four velocities of the initial and final particles \cite{c3}. It is the so called the self-similarity 
approach, which demonstrates a similarity of inclusive spectra of hadrons produced in $pp$ and $AA$ collisions, as a function
of similarity parameter $\Pi$. In fact, this approach is valid not in the complete kinematical region. That will be discussed in our paper.
The hadron inclusive spectra obtained within this approach are presented as a function of the relativistic invariant similarity parameter, 
which can be related to variables $t$ and $s$. The general form of such spectrum is not factorized over
$t$ and $s$. That is an advantage of the similarity approach in comparison to all the models mentioned above. However,
the unfactorized form of inclusive spectra is significant at not large initial energies and it becomes independent of $\sqrt{s}$
at large $\sqrt{s}$ like the ISR, SPS and LHC energies.
     
The approach of studying relativistic nuclear interactions in the four 
velocity space proved to be very fruitful \cite{c4}. In this article, we present a 
further development of this approach.

\section{The parameter of self-similarity.}
\label{se2}

Within the self-similarity approach \cite{c3,c4} the predictions on 
the ratios of particles produced in $AA$ collisions at high energies were given in \cite{c5}.
Let us briefly present here the main idea of this study. Consider, for example, 
the production of hadrons $1,2$, etc. in the collision of a nucleus $I$ with a nucleus II:
\begin{equation}
I  +   I\!I  \to  1  +  2  + \ldots                                           
\label{eq:n1}                                                                
\end{equation}                           
    
    According to this assumption more than one nucleon in the nucleus I can participate in 
the interaction (\ref{eq:n1}). The value of $N_I$ is the effective 
number of nucleons inside the nucleus $I$, participating in the interaction 
which is called the cumulative number. 
Its values lie in the region 
of $0 \le N_I \le A_I$ ($A_I$ - atomic number of nucleus I). The cumulative area complies with $N_I > 1$. Of course, 
the same situation will be for the nucleus $I\!I$, and one can enter the 
cumulative number of $N_{I\!I}$.

    For reaction (\ref{eq:n1}) with the production of the inclusive particle 1
we can write the conservation law of four-momentum in the following form: 
$${(N_IP_I + N_{I\!I}P_{I\!I} - p_1)}^2 = $$
\begin{equation}
{(N_Im_0 + N_{I\!I} m_0 + M)}^2 ,
\label{eq:n2}
\end{equation}

\noindent where $N_I$ and $N_{I\!I}$ 
the number of nucleons involved in the interaction; 
$P_I$, $P_{I\!I}$ , $p_1$
are four momenta of the nuclei $I$ and ${I\!I}$ and particle $1$, 
respectively; $m_0$
is the mass of the nucleon; $M$ is the mass of the particle 
providing the conservation of the baryon number, 
strangeness, and other quantum numbers.

    In \cite{c6} the parameter of self-similarity is introduced, which allows one to 
describe the differential cross section of the yield of a large class of 
particles in relativistic nuclear collisions:

\begin{equation} 
\Pi=\min[\frac{1}{2} [(u_I N_I + u_{I\!I} N_{I\!I})^2]^{1/2} ,
\label{eq:n3} 
\end{equation}

where $u_I$ and $u_{I\!I}$  are four velocities of the nuclei  $I$ and 
${I\!I}$. The values $N_I$ and $N_{II}$ will be measurable, if we accept the hypothesis of minimum
 mass $m_{0}^{2}(u_1N_1 +u_2N_2)^2$ and consider the conservation law of 4-momentum. Thus,
 the procedure to determine $N_I$ and $N_{II}$, and hence $\Pi$, is the determination of the
 minimum of $\Pi$ on the basis of the conservation laws of energy-momentum.

Then, it was suggested \cite{c5,c6} that the inclusive spectrum of the produced particle $1$ in $AA$ collision
can be presented as the universal function dependent of the self-similarity parameter $\Pi$, which was chosen,
for example, as the exponential function:

$$ E d^3 \sigma/dp^3 = $$
\begin{equation}
C_1 A_I^{\alpha(N_I)} \cdot 
A_{I\!I}^{\alpha(N_{I\!I})} \cdot \exp(-\Pi/C_2),
\label{eq:n4} 
\end{equation}

where $\alpha(N_I)=1/3 + N_I/3$, \\
$\alpha(N_{I\!I})=1/3 + N_{I\!I}/3$, \\
$C_1=1.9 \cdot 10^4 mb \cdot GeV^{-2} \cdot c^3 \cdot st^{-1}$ 
and \\
$C_2 = 0.125 \pm 0.002$.

\section{Self-similarity parameter in the central rapidity region}
\label{se3}

In the mid-rapidity region (y=0, y is the rapidity of particle 1) the analytical 
form for $\Pi$ was found in \cite{c5}. In this case $N_I$ and $N_{I\!I}$ 
are equal to each other: $N_I = N_{I\!I} = N$.
\begin{equation}
N = [1 + (1 + \Phi_{\delta}/\Phi^2)^{1/2}] \Phi ,
\label{eq:n5}
\end{equation}

\noindent where
\begin{equation}
\Phi = 2m_0 (m_t ch Y + M)/ sh^2 Y ,
\label{eq:n6}
\end{equation}

\begin{equation}
\Phi_{\delta}=(M^2 - m_1^2)/(4m_0^2 \cdot sh^2 Y) .
\label{eq:n7}
\end{equation}

     Here $m_{1t}$ is the transverse mass of the particle~1, 
$m_{1t} = (m_1^2 + p_t^2)^{1/2}$, $Y$- rapidity of interacting nuclei.

\noindent And then -
\begin{equation}
\Pi = N \cdot ch Y .
\label{eq:n8}
\end{equation}

\noindent This formula is obtained by searching for the minimum of the expression for $\Pi$ by differentiation
of the right-hand side of equation (3) on variables  $N_I$ and $N_{II}$  taking into account the conservation law (2).

\noindent For baryons we have
\begin{equation}
\Pi_b = (m_{1t} ch Y - m_1) ch Y/(m_0 sh^2 Y)                         
\label{eq:n10}
\end{equation}

\noindent and for antibaryons -
\begin{equation}                         
\Pi_a = (m_{1t} ch Y + m_1) ch Y/(m_0 sh^2 Y) .
\label{eq:n11}
\end{equation}

\begin{figure}[h]
\begin{center}
\includegraphics[width=0.8\textwidth]{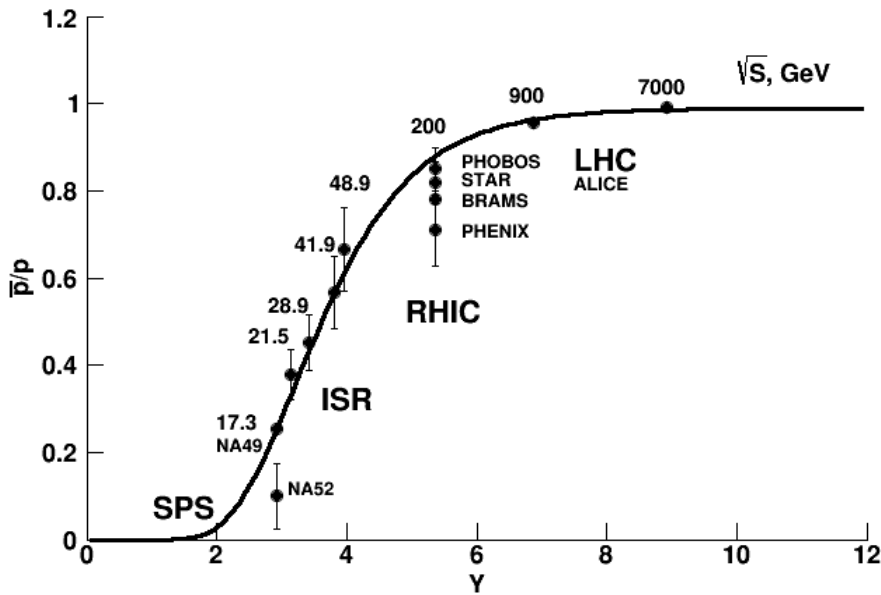} 
 \caption{\label{fig1}
The dependence of ratio of the antiproton cross section to the proton one as a function of initial rapidity
$Y$ or energy ($\sqrt{S}$, GeV) of the interacting nuclei.The points are the experimental data \cite{c7,c8,c9}.  
}
\end{center}
\end{figure}

   The results of calculations for the ratio of the antiproton cross section 
to the proton one after integration of Eqs.(\ref{eq:n10},\ref{eq:n11}) over $dm_{1t}$ 
are presented in 
Fig.~\ref{fig1}. This ratio is compared to the experimental data including the latest 
data at LHC \cite{c7,c8,c9}. 

     One can see a very good agreement of the results of our calculations with the 
experimental data in the wide energy range.

     The same calculations can be made for other antiparticles and particles. However, there are 
poor experimental data in the central rapidity region.

\section{Further development of Baldin's  approach.}
\label{se4}

As it is mentioned above, the exponential form for the hadron inclusive spectrum given by Eq.(\ref{eq:n4})
was chosen as an example and using it we can satisfactorily describe the ratio of total yields of 
antiprotons to protons produced in heavy nucleus-nucleus collisions. Unfortunately, this simple 
form Eq.(\ref{eq:n4}) contradicts to the LHC data on the inclusive spectra of hadrons produced in the 
central $pp$ collisions, as shown in \cite{c12}. Therefore, we use the results of \cite{c12} 
to present the inclusive relativistic invariant hadron spectrum at the mid-rapidity region and at 
not large hadron transverse momenta $p_{t}$ in a more complicated form, which 
consists of two parts. The first one is due to the contribution of quarks, which was obtained within
the QGSM \cite{ABK:1982,ABK:1999} using the AGK (Abramovsky, Gribov, Kanchelli) 
cancellation \cite{c10} of $n$-pomeron exchanges for inclusive hadron spectra at the mid-rapidity region.
It is written in the following form \cite{c12}: 

$$E(d^3 \sigma/d^3p)_q =
\phi_q(y=0,p_t) \cdot \sum_{n=1}^{\infty}[n \sigma_n(s)] =$$ 
\begin{equation}
\phi_q(y=0, p_t) g(s/s_0)^{\Delta}~,
\label{eq:n14}
\end{equation}
where $\sigma_n(s)$ is the cross-section for production of the $n$-pomeron chain (or 2n quark-antiquark strings);
$g=21$ mb - constant, which is calculated within 
the "quasi-eikonal" approximation \cite{c11};
$s_0 = 1~GeV^2$; $\Delta = [\alpha_p(0)-1] \sim 0.12$, where $\alpha_p(0)$ is the sub critical 
Pomeron intercept \cite{ABK:1982,ABK:1999,c11}.                     

The second part of the hadron inclusive spectrum at the mid-rapidity region was introduced in \cite{c12,c13}
assuming the contribution of the nonperturbative gluons and calculating it as the one-pomeron exchange between two
nonperturbative gluons in the collided protons \cite{c13}.
This part was written in the following form \cite{c12}:
$$E(d^3 \sigma/d^3p)_g =
\phi_g(y=0,p_t) \cdot  \sum_{n=2}^{\infty}(n-1) \sigma_n(s) =$$
\begin{equation}
\phi_g(y=0,p_t) \cdot \left(\sum_{n=1}^\infty n\sigma_n(s)- \sum_{n=1}^\infty \sigma_n(s)\right)=
\phi_g(y=0,p_t) \cdot [g(s/s_0)^{\Delta} - \sigma_{nd}]~,
\label{eq:n15}
\end{equation}
where $\sigma_{nd}$ is the non diffractive $pp$ cross section.

Thus, taking into account the quark and gluon contributions we will 
get the following form for the inclusive hadron spectrum:
\begin{equation}
E(d^3 \sigma/d^3p)~=~[\phi_q(y=0,p_t) + \phi_g(y=0,p_t) \cdot 
(1 - \sigma_{nd}/g((s/s_0)^{\Delta})]
\cdot g(s/s_0)^{\Delta}
\label{eq:n16}
\end{equation}

The question arises, what is a relation of the similarity parameter $\Pi$ to the  
relativistic invariant variables $s,p_t^2$ ? 
This relation can be found from Eqs.(\ref{eq:n5}-\ref{eq:n8}) using $ch(Y)=\sqrt{s}/(2m_0)$.
Then, we have the following form for $\Pi$:                                                                                                                                                                                                                                                                                                                                                                                                                                                                                                                                                                                                                                                                                                                                                                                                                                                                                                                                                                                                                                                                  
\begin{equation}
\Pi=\left\{\frac{m_{1t}}{2m_0\delta}+\frac{1}{\sqrt{s}\delta}\right\}
\left\{1+\sqrt{1+\frac{M^2-m_1^2}{m_{1t}^2}\delta}\right\}
\label{def:Pi}
\end{equation}
where $\delta=1-4m_0^2/s$; $m_{1t}=\sqrt{p_{t}^2+m_1^2}$ is the transverse mass of the produced hadron $h$.
At large initial energies $\sqrt{s}>>$ 1 GeV the similarity parameter $\Pi$ becomes 
\begin{equation}
\Pi=\frac{m_{1t}}{2m_0(1-4m_0^2/s)}
\left\{1+\sqrt{1+\frac{M^2-m_1^2}{m_{1t}^2}(1-4m_0^2/s)}\right\}
\label{def:Picorr}
\end{equation}
For $\pi$-mesons $m_1=\mu_\pi$ is the pion mass and $M=0$; for $K^-$-mesons $m_1=m_K$ is the kaon mass and
$M=m_K$; for $K^+$-mesons $m_1=m_K$ and $M=m_\Lambda - m_K$, $m_\Lambda$ is the mass of the $\Lambda$-baryon.   
For $\pi$-mesons at $p_t^2>>m_1^2$ we have: 
\begin{equation}
\Pi\simeq \frac{m_{1t}}{m_0(1-4m_0^2/s)} 
\label{def:Piappr}
\end{equation}
One can see that in a general case the similarity parameter $\Pi$ 
depends on $p_t^2$ and $s$ and asymptotically at large $s>>4m_0^2$ it depends only on $p_{t\pi}^2$. Let us
stress that the dependence of $\Pi$ on $s$ is crucial at low initial energies only.
 
The invariant inclusive spectrum can be also presented in the following equivalent form: 
\begin{equation}
E\frac{d^3 \sigma}{d^3p}~=~\frac{1}{\pi}\frac{d\sigma}{dp_t^2dy}\equiv \frac{1}{\pi}\frac{d\sigma}{dm_{1t}^2dy}
\label{def:spypt}
\end{equation} 

Taking into account (\ref{def:spypt}) we can rewrite Eq.(\ref{eq:n16}) in the form:
\begin{equation}
\frac{1}{\pi}\frac{d\sigma}{dm_{1t}^2dy}=
[\phi_q(y=0,\Pi) + \phi_g(y=0,\Pi) \cdot(1 - \sigma_{nd}/g((s/s_0)^{\Delta})]
\cdot g(s/s_0)^{\Delta}~.
\label{eq:n21}
\end{equation}
 The first part of the 
inclusive spectrum (Soft QCD (quarks)) is related to the function $\phi_q(y=0,\Pi)$, which is fitted by the following form \cite{c12}:
\begin{equation}
\phi_q(y=0,\Pi)~=~A_q exp(-\Pi/C_q)~,
\label{eq:fiqfit}
\end{equation} 
where $A_q=3.68~(GeV/c)^{-2}, C_q=0.147~GeV/c^2$.

The function $\phi_g(y=0,\Pi)$ related to the second part (Soft QCD (gluons)) of the spectrum is fitted by the following
form \cite{c12}:
\begin{equation}
\phi_g(y=0,\Pi)~=~A_g\sqrt{m_{1t}} exp(-\Pi/C_g)~,
\label{eq:figfit}
\end{equation}
where $A_g=1.7249~(GeV/c)^{-2}, C_g=0.289~GeV/c^2$.     
\begin{figure}[h!] 
\begin{center}
\epsfig{file=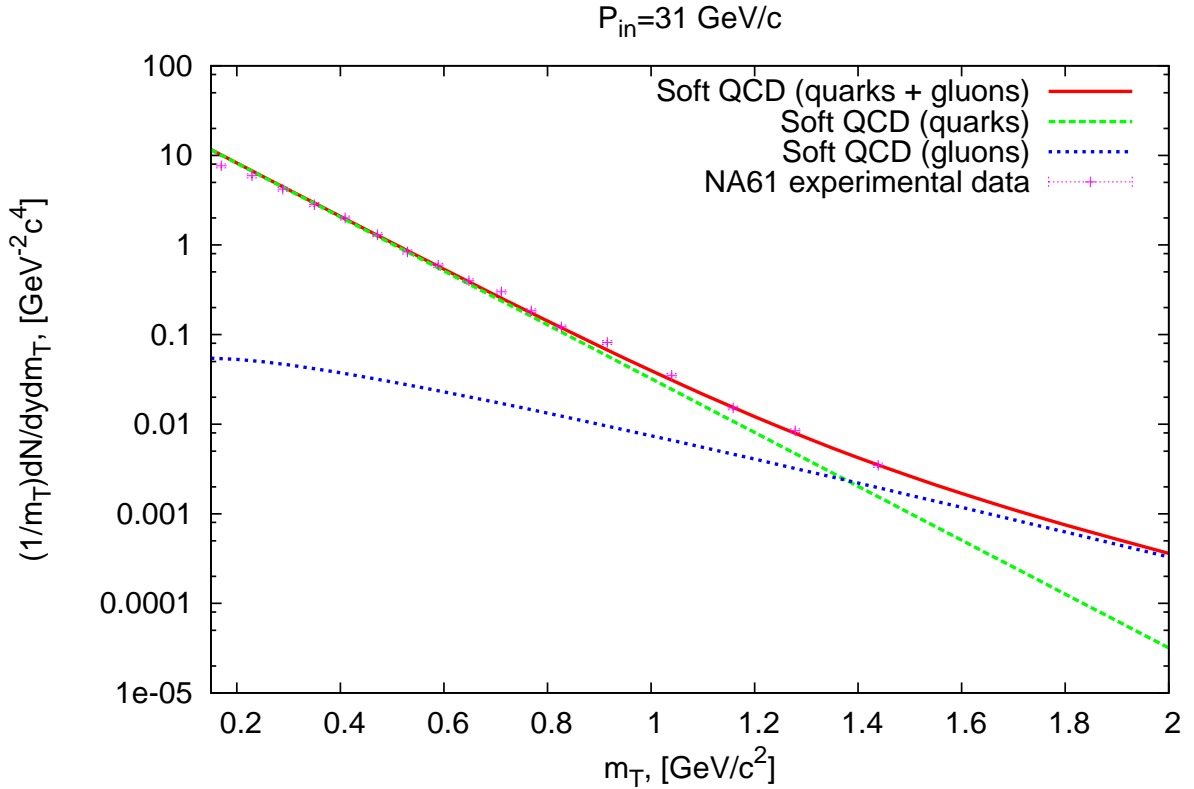,width=1.00\linewidth}
\end{center}
 \caption{
Results of the calculations of the inclusive cross section of hadron 
production in $pp$ collisions as a function of the transverse mass at the initial momenta
$P_{in}=31~GeV/c$. They are compared to the NA61
experimental data \cite{c14}.
} 
\label{fig2}
\end{figure}

\begin{figure}[h!] 
\begin{center}
\epsfig{file=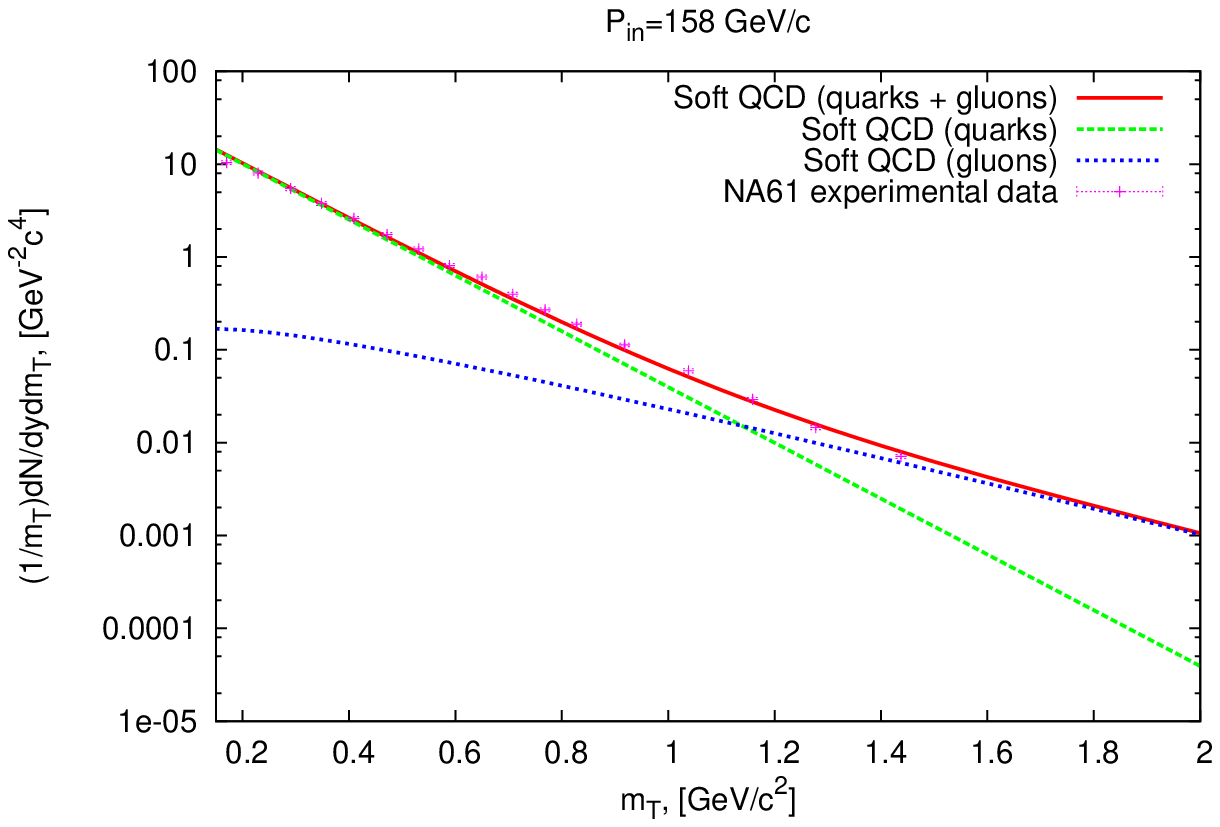,width=1.00\linewidth}
\end{center}
 \caption{
Results of the calculations of the inclusive cross section of hadron 
production in $pp$ collisions as a function of the transverse mass at the initial momenta
$P_{in}=158~GeV/c$. They are compared to the NA61
experimental data \cite{c14}.
} 
\label{fig3}
\end{figure}


     Using (\ref{eq:n21}) we can calculate the inclusive hadron spectrum 
as a function of the transverse mass.                                  
    
    In Figs.~\ref{fig2}, \ref{fig3} the inclusive spectra
$(1/m_{1t})d\sigma/dm_{1t}dy$ of $\pi^-$-mesons produced in $pp$ collisions at the initial momenta $P_{in}=$ 31 GeV$/$c and 
$P_{in}=$ 158~GeV$/$c are presented versus their transverse mass $m_t$. 
Using only the first part of the spectrum $\phi_q(y=0, m_{1t})$, which is due  to
the quark contribution, the conventional string model, let's call it the SOFT QCD (quarks), one can describe the NA61 data \cite{c14}
rather satisfactorily at $p_{in}=$31~GeV$/$c and $m_t<$ 1~GeV$/$c$^2$ . This part of the inclusive spectrum 
corresponds to the dashed line in Fig.~\ref{fig2}.
The inclusion of the second part of spectrum due to the contribution of gluons (SOFT QCD (gluons)), the dotted line, allowed 
us to describe all the NA61 data up to $m_t~$ 1.5 Gev$/$c, see the solid line in Fig.~\ref{fig2} (Soft QCD(quarks+gluons)). 
Actually, at large $\sqrt{s}$ even at the NA61 energies $\Pi\simeq m_{1t}/m_0$ instead of ~(\ref{def:Piappr}).
Generally the pion spectrum $\rho(s,m_{1t}\equiv E(d^3 \sigma/d^3p)$
(ignoring the gluon part) can be presented in the following 
approximated form, which is valid for the NA61 energies and low transverse momenta $p_t<1~GeV/c$:
\begin{eqnarray}
\rho(s,m_{1t})~\simeq~\phi_q(y=0,\Pi)g(s/s_0)^{\Delta}=g(s/s_0)^{\Delta}A_q exp(-\frac{m_{1t}}{C_qm_0(1-4m_0^2/s)})\\
\nonumber
\equiv g(s/s_0)^{\Delta}A_q exp(-\frac{m_{1t}}{T})~, 
\label{def:rhoq_appr}
\end{eqnarray}
where 
\begin{eqnarray}
T=C_qm_0(1-4m_0^2/s)
\label{def:Ts} 
\end{eqnarray}
is the inverse slope parameter, which is called sometime as the {\it thermal freeze-out} temperature.
One can see from Eq.(\ref{def:Ts}) that this {\it thermal freeze-out} temperature depends on the initial energy square $s$ in the c.m.s.
of collided protons. That is the direct consequence of the self-similarity approach, which operates the four-momentum velocity formalism. 
This $s$-dependence of $T$ is significant at low initial energies and at $s>>m_0^2$ the slope parameter $T$ becomes independent of $s$.
To describe rather well the NA61 data at larger values of $p_t$, the inclusive pion spectrum should be presented 
by Eq.(\ref{eq:n21}), which has a more complicated form in comparison to the simple exponential one of 
(23). However, the main contribution to the inelastic total cross section comes from the
first part of Eq.(\ref{eq:n21}), which has the form given by Eq.(21)  

\begin{figure}[h!]
\begin{center}
\includegraphics[width=0.8\textwidth]{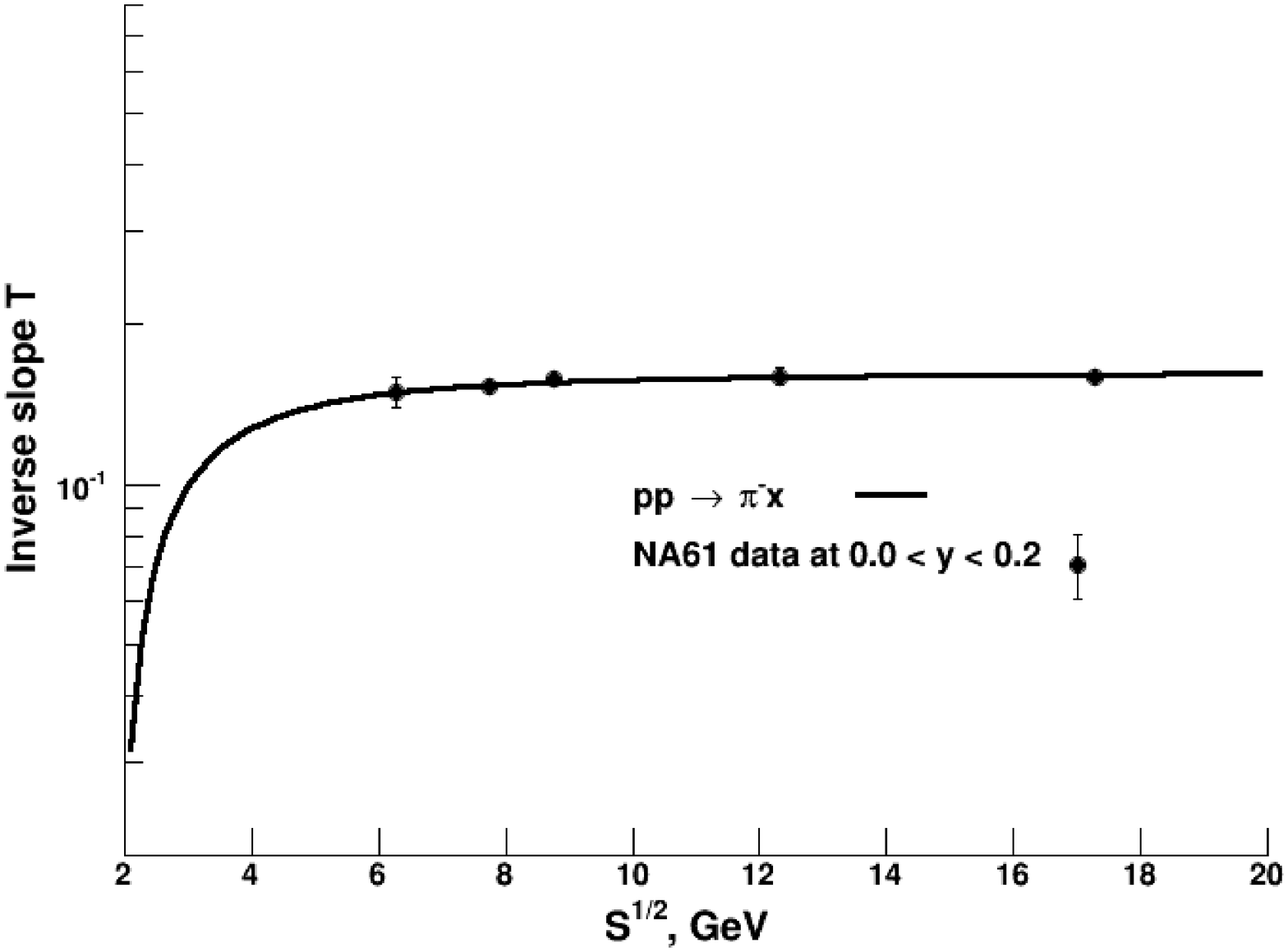} 
 \caption{
Results of calculations of the inverse slope parameter 
$T$ on the energy dependence for the negative pion production in 
pp-interactions. The experimental points are taken from \cite{c14}.}
\end{center}
\label{fig_slope}
\end{figure}

We have calculated the inverse slope parameter T, as a function of the energy $\sqrt{s}$ given by Eq.(\ref{def:Ts}) and presented in 
Fig.~\ref{fig_slope}. There is a good agreement with the experimental data \cite{c14}. 
     
\begin{figure}[h!] 
\begin{center}
\epsfig{file=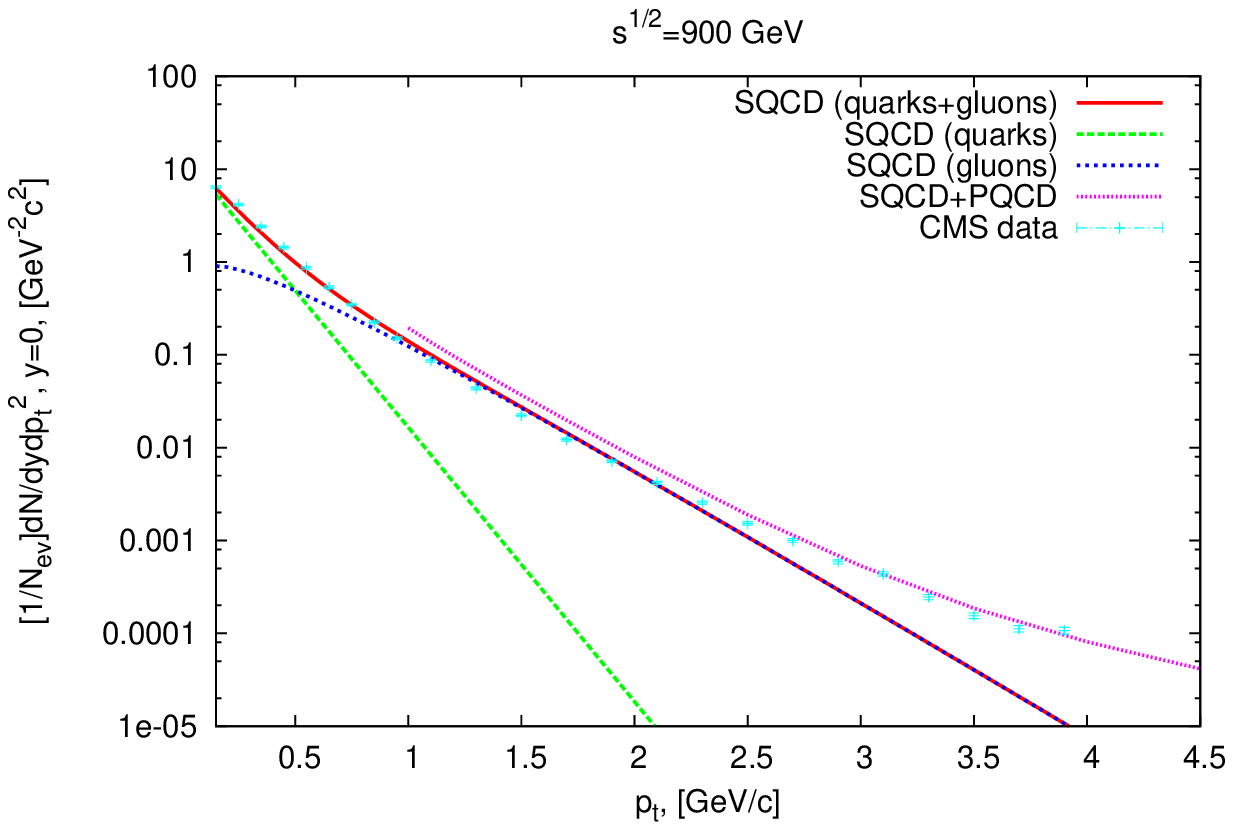,width=1.00\linewidth}
\end{center}
 \caption{
Results of the calculations of the inclusive cross section of charge hadrons
produced in $pp$ collisions at the LHC energies as a function of their transverse momentum $p_t$
at $\sqrt{s}=$0.9~TeV. The points are the LHC experimental data \cite{CMS}.
}  
\label{fig5}
\end{figure}

\begin{figure}[h!] 
\begin{center}
\epsfig{file=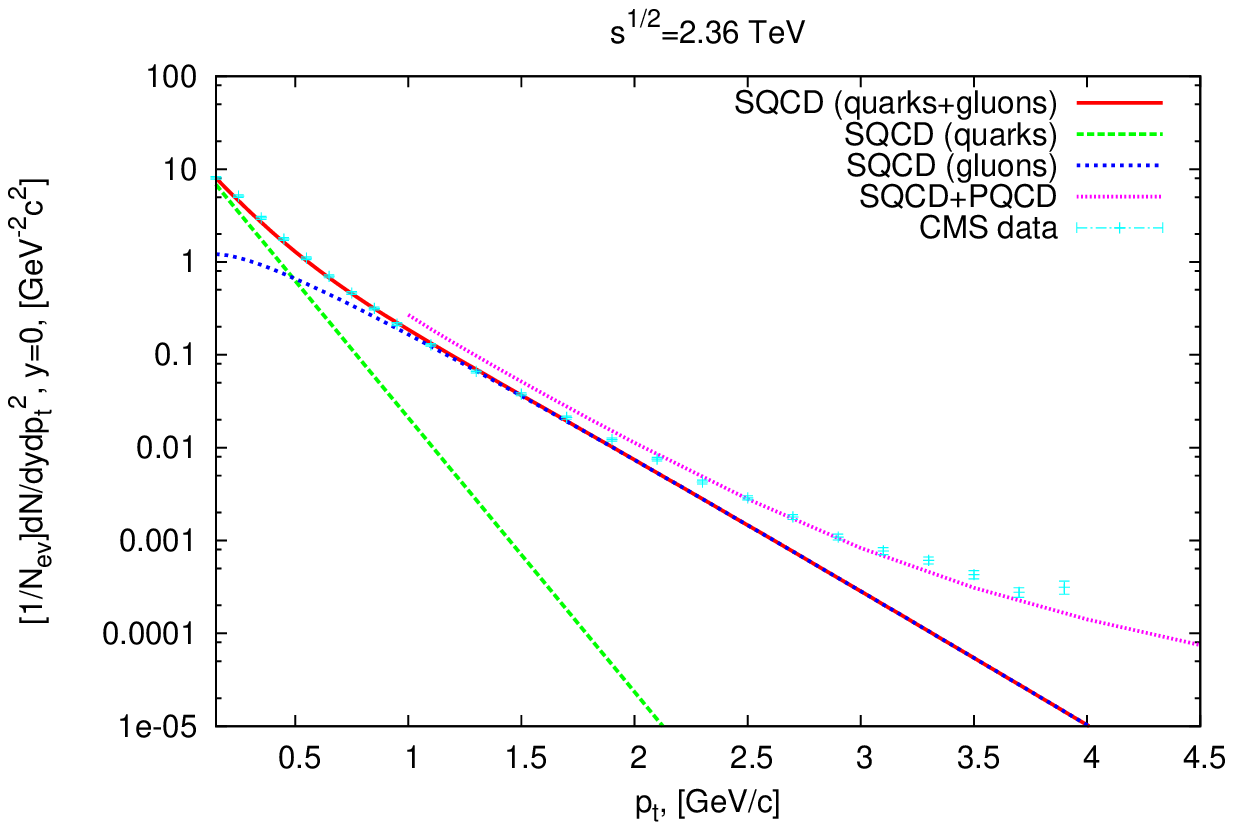,width=1.00\linewidth}
\end{center}
 \caption{
Results of the calculations of the inclusive cross section of charge hadrons
produced in $pp$ collisions at the LHC energies as a function of their transverse momentum $p_t$
at $\sqrt{s}=$2.36~TeV. The points are the LHC experimental data \cite{CMS}.
}  
\label{fig6}
\end{figure}

\begin{figure}[h!] 
\begin{center}
\epsfig{file=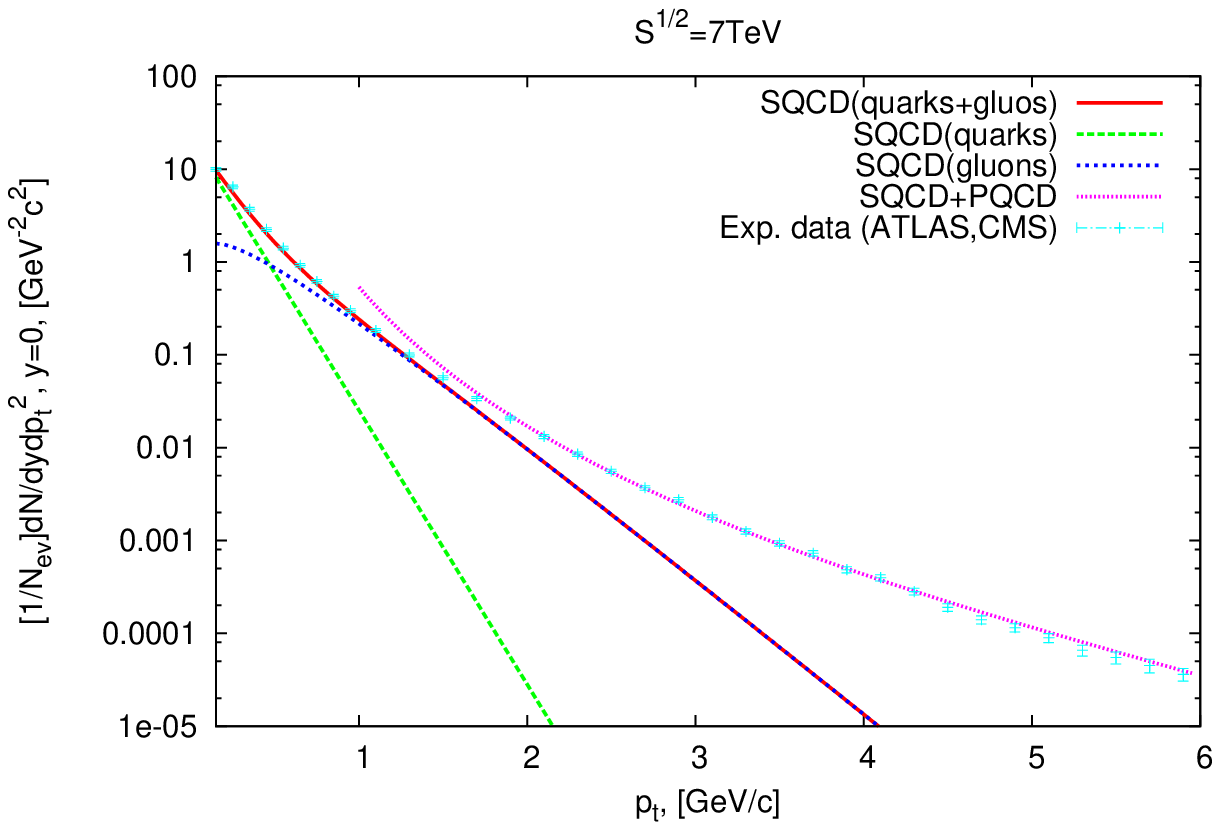,width=1.00\linewidth}
\end{center}
 \caption{
Results of the calculations of the inclusive cross section of charge hadrons
produced in $pp$ collisions at the LHC energies as a function of their transverse momentum $p_t$
at $\sqrt{s}=$7~TeV. The points are the LHC experimental data \cite{CMS, ATLAS}.
}  
\label{fig7}
\end{figure}

In Figs.~\ref{fig5}-\ref{fig7} we give the calculations of inclusive spectra of charged 
hadrons (mainly pions and kaons) produced in $pp$ collision at $\sqrt{s}=$ 900 GeV, 2.36 TeV, 7 TeV performed 
by using (\ref{eq:n16}) and the perturbative QCD (PQCD) within the LO \cite{c12,c13} compared to the LHC data.
These spectra are the sums of inclusive spectra of pions and kaons, therefore they are presented as a function of 
the transverse momentum $p_t$ instead of functions of the transverse mass $m_{1t}$ because the masses of a pion and
kaon are different.    
In addition to the part of spectrum, which corresponds to (\ref{eq:n16}), see the solid lines in these figures, 
we also include the PQCD calculations, see the dotted lines. The PQCD calculation within the LO is 
divergent at low $p_t$, therefore, the dotted lines go up, when $p_t$ decreases. The kinematical region 
about $p_t\simeq$ 1.8-2.2 GeV$/$c$^2$ can be treated as the matching region of the nonpertubative QCD (soft QCD) and
the pertubative QCD (PQCD). One can see from Figs.~\ref{fig5} - \ref{fig7} that it is possible to describe rather well
these inclusive spectra in the wide region of $p_t$ at the LHC energies matching these two approaches. 
Figs.~\ref{fig2},\ref{fig3},\ref{fig5}-\ref{fig7} show that in order to describe the inclusive hadron spectra \cite{CMS,ATLAS}
in the mead-rapidity region at $m_{1t}<$ 2 GeV$/$c$^2$, the form of spectra presented in a simple exponential form 
of (\ref{eq:n14}) should be changed and can be presented in the form of (\ref{eq:n16}), which includes the 
non trivial energy dependence.
To describe rather well the LHC data on these inclusive $p_t$-spectra at $p_t>$ 2-3 GeV$/$c, the PQCD calculation should be 
included, the contribution of which has a shape similar to the power law $p_t$-distribution \cite{CLPST:2014}.

\section{Conclusion}
  
The inclusive hadron spectrum in the space of four-velocities is presented within the self-similarity
approach as a function of the similarity parameter $\Pi$.
The use of the self-similarity approach allows us to describe the ratio of the total yields
of protons to anti-protons produced in A-A collisions as a function of the energy in the mid-rapidity region
and a wide energy range from 10~GeV to a few TeV. To study the similar ratio of light nuclei to 
anti-nuclei, we need more detailed experimental data.

We have shown that the energy dependence of the similarity parameter $\Pi$ included within this approach is very significant at low energies,
namely at $\sqrt{s}<$ 6~GeV, and rather well reproduces the experimental data on the inverse slope or the {\it thermal freeze-out}
temperature of the inclusive spectrum of hadrons produced in $pp$ collisions, it increases and saturates when $\sqrt{s}$ grows. 
This is very significant for a theoretical interpretation 
of future experimental data planned to get at FAIR,~CBM (Darmstadt, Germany) and NICA (Dubna, Russia) projects. That is 
an advantage of the self-similarity approach compared to other theoretical models.

However, we have also shown that the $s$ dependence of $\Pi$ is not enough to describe
the inclusive spectra of hadrons produced in the mid-rapidity region, for example, in $pp$ collisions in the wide region of
initial energy, especially at the LHC energies. Therefore, we modify the self-similarity approach using the quark-gluon string model 
(QGSM) \cite{ABK:1982,ABK:1999} and \cite{c12,c13} including the contribution of nonperturbative gluons, which are very
significant to describe the experimental data on inclusive hadron spectra in the mid-rapidity region at the transverse momenta 
$p_t$ up to 2-3~GeV$/$c \cite{c12,c13}. Moreover, the gluon density obtained in \cite{c13}, parameters of which were found from the best 
description of the LHC data, allowed us also to describe the HERA data on the proton structure functions \cite{c15}. To describe the data 
in the mid-rapidity region and values of $p_t$ up to 2-3~GeV$/$c, we modify the simple exponential form of the spectrum, as a function of $\Pi$,  
and present it in two parts due to the contribution of quarks and gluons, each of them has different energy dependence. This energy dependence
was obtained in \cite{c12} based on the Regge approach valid for soft hadron-nucleon processes. 
To extend the application of the suggested approach to analyze these inclusive $p_t$-spectra at large hadron transverse momenta, we have to 
include the PQCD calculation, which results in the main contribution at $p_t>$ 2-3 GeV$/$c.

\end{document}